\documentclass[twocolumn,showpacs,aps,prl,superscriptaddress]{revtex4}

\usepackage{graphicx}

\begin{document}
\title{Extended quantum critical phase in a magnetized spin-$\frac{1}{2}$
antiferromagnetic chain}

\author{M. B. Stone}
 \altaffiliation[Currently at ]{Department of Physics, The
Pennsylvania State University, University Park, Pennsylvania
16802}
\affiliation{Department of
Physics and Astronomy,
Johns Hopkins University, Baltimore, Maryland 21218}
\author{D. H. Reich}
\affiliation{Department of
Physics and Astronomy,
Johns Hopkins University, Baltimore, Maryland 21218}
\author{C. Broholm}
\affiliation{Department of
Physics and Astronomy,
Johns Hopkins University, Baltimore, Maryland 21218}
\affiliation{National Institute of Standards and
Technology, Gaithersburg, MD 20899} 

\author{K. Lefmann}
\affiliation{Materials Research Department, Ris\o\ 
National Laboratory, DK-4000 Roskilde, Denmark}

\author{C. Rischel}
\affiliation{\O rsted Laboratory, Niels Bohr Institute,
University of Copenhagen, DK-2100, K\o benhavn
\O, Denmark}

\author{C. P. Landee}
\affiliation{Carlson School of Chemistry and Department of
Physics, Clark University, Worcester, MA 01610}
\author{M. M. Turnbull}
\affiliation{Carlson School of Chemistry and Department of
Physics, Clark University, Worcester, MA 01610}

\date{\today}

\begin{abstract}
Measurements are reported of the magnetic field dependence
of excitations in the
quantum critical state of  the spin $S=1/2$ linear chain
Heisenberg antiferromagnet   copper pyrazine
dinitrate (CuPzN).  The complete spectrum was measured
at $k_{B}T/J \le 0.025$ for $H=0$ and $H=8.7$ Tesla  where
the system is $\sim 30$\% magnetized. At $H=0$, the results
are in quantitative agreement with exact calculations
of the dynamic spin correlation function for a two-spinon
continuum.
At high magnetic field, there are multiple overlapping continua with 
incommensurate soft modes.  
The boundaries of
these continua confirm long-standing predictions, and
the intensities are consistent with exact diagonalization and Bethe Ansatz calculations.

\end{abstract}

\maketitle

One of the most important ideas to emerge from studies 
of condensed matter systems in recent years is the concept of
quantum criticality \cite{sachdevbook}.   A quantum critical point marks a zero
temperature phase transition between different ground states
of a many-body system as a result of changes in parameters
of the underlying Hamiltonian.  Precisely at the quantum critical point
the system is without characteristic length scales or energy scales, with power-law
spatial correlations  and gapless excitations. Finite temperature properties close to 
quantum criticality are anomalous and reflect universal properties of the underlying 
quantum field theory. 

While fine tuning of a parameter in a system's
Hamiltonian is generally required to achieve quantum criticality, 
it is inherent to the spin S=1/2 linear chain
Heisenberg antiferromagnet (LCHAFM).  The spin dynamics of the LCHAFM
have been studied in a number of materials
\cite{Tennant95,dendercubenzprb,takigawa9697,araiprl1996,hammarprb1999},   
and the elementary excitations are S=1/2 spinons that form a gapless, two-particle 
continuum \cite{muller1981,karbach1997}. 
In a magnetic field, $H$ the Hamiltonian of the LCHAFM is $
\mathcal{H}\mathit{=\sum_{i}[ J}
\mathbf{S}\mathit{_{i}}\mathbf{S}\mathit{_{i+1}-g\mu_{B}}H
\mathrm{S}_{i}^{z}\mathit{]}$.
While changes in ${\mathcal H}$ such as the introduction of
dimerization or interchain coupling drive the LCHAFM away from criticality
\cite{chitraprb1997}, the system should remain quantum critical at $T=0$ in fields below 
the fully spin polarized state, which occurs at $H_C = 2J/g\mu_B$ 
\cite{bogoliubov1986,fledderjohann1997}.
Prominent among the features  predicted to exist along this quantum critical 
line are a set of 
field-dependent two-particle continuua 
\cite{muller1981,lefmannprb1996,karbach_prb_2000,Karbach02} 
with incommensurate soft modes that move  across the 1D Brillouin zone with increasing
field as $\tilde{q}_{i,1} = 2 \pi m$ and 
$\tilde{q}_{i,2} = \pi - 2 \pi m$, where $0 \le m \le 1/2$ is the
reduced magnetization per spin.

Experiments  on copper benzoate \cite{dendercubenzprl}
have verified the predicted
field-dependence of the incommensurate wavevector $\tilde{q}_{i,2}$. However, 
due to a combination of a staggered $g$-tensor and Dzayaloshinskii-Moria
interactions, the field drives that system away from the critical line to a state with 
confined spinons and a gap in the excitation spectrum 
\cite{AffleckOshikawa}.
In contrast, the finite-field critical state of the  S=1/2  LCHAFM is  accessible in 
copper pyrazine dinitrate, $\rm Cu(C_{4}H_{4}N_{2})(NO_{3})_{2}$ (CuPzN).  
This well characterized organo-metallic magnet has
$J = 0.9 $ meV, and  negligible interchain
coupling ($J^{\prime}/J < 10^{-4})$ \cite{hammarprb1999,losee1973}. 
The spin chains in CuPzN are uniform, with one 
Cu$^{2+}$ ion per unit cell along the chain, and
specific heat measurements have shown that CuPzN remains gapless for
 $H \le 0.6 H_C = 9$ T \cite{hammarprb1999}.
In this paper we report inelastic neutron scattering measurements
of the full spectrum of CuPzN, both at $H=0$ and at 
$H=8.7$ T, where $m = 0.15$.
In zero field, the spectrum is consistent with the
exact two-spinon contribution to the spin
fluctuation spectrum 
\cite{karbach1997}.
The finite field data show the
long-sought
field-dependent continua \cite{muller1981}, and are in
detailed agreement with theoretical and numerical work 
\cite{lefmannprb1996,karbach_prb_2000,Karbach02}.

The measurements were performed using the
SPINS cold neutron triple axis spectrometer at the NIST Center for Neutron Research. 
The sample studied  contained  22
single crystals of deuterated CuPzN \cite{StoneThesis} with a total  mass of 3.06
grams,and a measured mosaic of two degrees.
We used a dispersive analyzer configuration that detects scattered neutrons with an 
angular range of 10$^{\circ}$ and an energy range 
2.7 meV $\le E_{f}\le$ 3.55 meV \cite{zaliznyak}. We used
a Be filter before the sample for 
$E_i < 5.15$ meV, and a BeO filter after the 
sample.  The measured full width at half maximum (FWHM)
elastic energy resolution was $\delta\hbar\omega = 0.14$ meV, and the
 FWHM wave-vector resolution along the [100] chain axis
was $\delta{\mathbf{Q}}_{\parallel}=0.03$ \AA$^{-1}$ at
${\mathbf{Q}}= (hkl) = (\frac{3}{2}\frac{1}{4}0)$ at $\hbar\omega=0$
\cite{chesser1973}. Data were obtained along
the line between ${\mathbf{Q}} =(1\frac{1}{4}0)$ and
${\mathbf{Q}}=(2\frac{1}{4}0)$, with $H \parallel \hat{c}$.   

\begin{figure}
\includegraphics[scale=0.5,bb=0 350 400 792,clip=true]{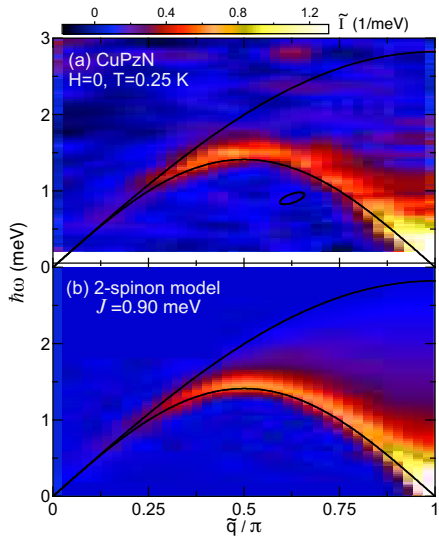}
    \caption{(a) Magnetic inelastic neutron scattering intensity $\tilde{{\mathcal I}}_{m}$
    for CuPzN at $T=0.25$ K and $H=0$
    versus wave-vector and energy transfer.  
    (b) Calculated two-spinon contribution to $\tilde{{\mathcal I}}_{m}$.  
    Solid lines are the predicted
    lower and upper bounds of the spinon continuum.  A
    representative FWHM resolution ellipsoid is shown in (a).
    }
    \label{fig:zerofieldcontour}
\end{figure}

Figure \ref{fig:zerofieldcontour}(a)  shows the normalized magnetic
scattering intensity $\tilde{{\mathcal I}}_{m}(\tilde{q},\omega)$ 
for CuPzN, measured in zero field
at $T= 0.25$ K.  Wave-vector
transfer along the chain is represented as $\tilde{q}=2\pi(h-1)$,
This data set was obtained by combining data
taken at $E_i = $  3.35 meV, 3.55 meV, and for 3.75 meV $\le E_i \le$ 5.75 meV
with 40 $\mu$V steps.  
After subtracting the background scattering measured with the 
 analyzer in a non-reflecting geometry,
 the incoherent elastic scattering profile of the sample was
 determined from 
 data that were at least 0.2 meV outside of the known bounds \cite{hammarprb1999}
of the $H=0$ spinon continuum. 
  This profile was 
scaled to to the measured elastic incoherent scattering intensity
at each $\tilde{q}$, and  subtracted from the raw data.
The data were converted to the normalized scattering
intensity  by comparison
with the measured incoherent scattering intensity of a vanadium
standard. 
The data at $2\pi - \tilde{q}$ were then
averaged with that at $\tilde{q}$,
 binned in bins of size $\delta\hbar\omega = 25$ $\mu$eV by
$\delta\tilde{q} = 0.026\pi$, and  averaged
 over a rectangle of the same size as the
FWHM energy and wave-vector resolutions to produce
the color contour plot shown in 
Fig.~\ref{fig:zerofieldcontour}(a).

\begin{figure}
\includegraphics[scale=0.45,bb=0 325 400 792,clip=true]{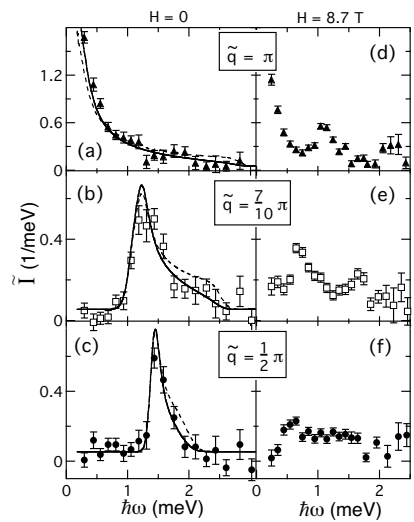}
   \caption{Magnetic scattering intensity vs $\hbar\omega$ at constant $\tilde{q}$
     for CuPzN at $T=0.25$ K and $H=0$  (a)-(c), and at
     $H=8.7$ T (d)-(f).  
     Each point includes scattering within a 
    width of $\Delta\tilde{q}=0.05 \pi$.  The solid (dashed)
     lines in (a)-(c) are 
   the exact  (approximate) model for $ {\mathcal S}(\tilde{q},\omega)$
   described in the text.    }
    \label{fig:constQ}

\end{figure}

\begin{figure}
\includegraphics[scale=0.45,bb=0 325 400 792,clip=true]{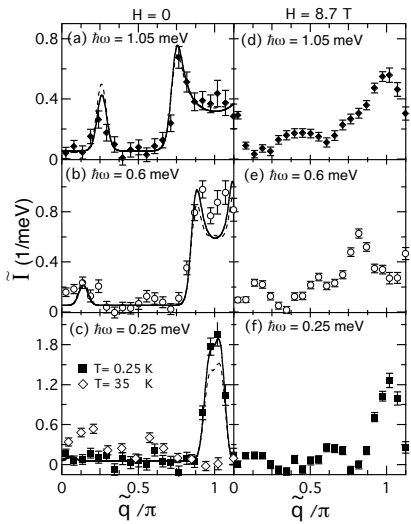}
    \caption{Magnetic scattering intensity vs $\tilde{q}$ at constant $\hbar\omega$
     for CuPzN at $T=0.25$ K at $H=0$  (a)-(c), and at
     $H=8.7$ T (d)-(f).  The data for $\tilde{q}  < \pi$ have
     been averaged with the data at $2\pi - \tilde{q}$.
     The data displayed for $\tilde{q} > \pi$
    have not been averaged in this manner.   Each point
    includes scattering within a 
    width of $\Delta\hbar\omega=0.1$ meV.  The solid (dashed)
     lines in (a)-(c) are 
   the exact  (approximate) model for $ {\mathcal S}(\tilde{q},\omega)$
   described in the text.  The scattering measured at $T=35$ K is
   shown as open symbols in panel (c).}
    \label{fig:constE}
\end{figure}

These data provide a complete picture
of the zero-field spinon continuum (SC) in CuPzN.
The solid lines 
are the SC's lower and upper bounds, calculated for 
$J = 0.9$ meV \cite{muller1981}. 
Figures \ref{fig:constE}(a)-(c) and \ref{fig:constQ}(a)-(c)
show cuts through the data, and highlight
several important features.
These include the
monotonically decreasing intensity at
$\tilde{q} = \pi$ [Fig.~\ref{fig:constQ}(a)],
the  peaks near $\tilde{q} = \pi$ associated
with the divergence in $ {\mathcal S}(\tilde{q},\omega)$ at the SC lower bound 
\cite{karbach1997} [Fig.\ref{fig:constE}(a)-(c)],
and the smaller feature at low $\tilde{q}$ where
the SC narrows as $\tilde{q} \rightarrow 0$.
Figures \ref{fig:constQ}(b)-(c) 
show the  asymmetric lineshapes produced by the  SC
for $\tilde{q} \ne \pi$.
Figure \ref{fig:constE}(c)
also shows data taken at $T=35$ K. The decrease in intensity at $\tilde{q} = \pi$
confirms that the low-$T$ scattering is magnetic. 

Figure \ref{fig:zerofieldcontour}(b) shows 
$\tilde{{\mathcal I}}_{m}(\tilde{q},\omega)$ calculated \cite{hammarprb1999}
from the recently derived exact two-spinon
contribution to  $ {\mathcal S}(\tilde{q},\omega)$
  \cite{bougourzi1996,fledderjohann1996,karbach1997}
with $J = 0.9$ meV,  with the addition of
a constant intensity fixed at the average residual background
measured 0.5 meV away from the continuum.
$\tilde{{\mathcal I}}_{m}(\tilde{q},\omega)$ is also
shown as solid lines on the cuts in Figs. \ref{fig:constE}(a)-(c) and 
\ref{fig:constQ}(a)-(c). There are no
adjustable parameters in the model, and
the agreement with the data is excellent,
 although we note
that there is an overall 10\% uncertainty in the
vanadium normalization. 

Figures \ref{fig:constE} and \ref{fig:constQ} also show a comparison (dashed lines)
to the lineshapes
produced by a global fit to the approximate form \cite{muller1981}
for $ {\mathcal S}(\tilde{q},\omega)$ that has been previously used to model
measurements of the $H=0$ SC
in CuPzN \cite{hammarprb1999} and other materials \cite{Tennant95,dendercubenzprb}.
This approximation is now known to overestimate
$ {\mathcal S}(\tilde{q},\omega)$ near the upper SC boundary 
\cite{karbach1997,lefmannprb1996}.
Indeed, some indications of this effect are
seen in Figs.~\ref{fig:constQ}(b) and  \ref{fig:constQ}(b),
and although our data are not optimal for observing
these differences, the exact model does give an overall
better description of the data.

\begin{figure}
\includegraphics[scale=2.1]{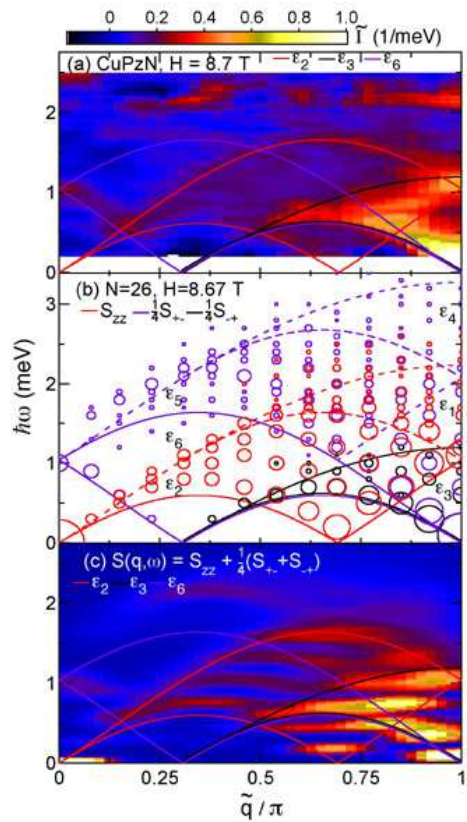}
    \caption{(a) Inelastic neutron scattering intensity 
    $\tilde{{\mathcal I}}_{m}(\tilde{q},\omega)$ for CuPzN at
    $T=0.25$ K
    and $H=8.7$ T.  (b) Calculations of the different
    components of $ {\mathcal S}(\tilde{q},\omega)$ for $N=26$ spins
    and $m= 2/13$ ($H=8.67$ T for CuPzN).
    The area of each circle is
    proportional to $ {\mathcal S}(\tilde{q},\omega)$.  
    (c)  $\tilde{{\mathcal I}}_{m}(\tilde{q},\omega)$
    calculated for ensemble of chains with $N=24$, $26$, and $28$.  Lines
    show bounds of excitation continua  for the
    finite-length $S=1/2$ LCHAFM.    Solid 
    lines: continua
    predicted to predominate  as $N \rightarrow \infty$.}
    \label{fig:fieldcontour}
 \end{figure}

The spectrum of CuPzN in a magnetic field is considerably
more complex than at zero field, as seen in 
Fig.~\ref{fig:fieldcontour}(a), which shows $\tilde{{\mathcal I}}_{m}$
 measured at $H = 8.7$ T.
This data set combines data taken in the range 3.35 meV $\le E_i \le$ 4.75 meV
with 20 $\mu$V increments, and at $E_i = 5.15$ meV.
An identical background subtraction, averaging,  and binning
procedure as used at $H=0$ was applied
to these data, using the $H=0$ incoherent elastic profile. 

Several continua  and prominent  features
have emerged that were not present at $H=0$.
These are  highlighted in cuts
 shown in Figs.~\ref{fig:constE}(d)-(f)
and \ref{fig:constQ}(d)-(f).  First,
a continuum of scattering is still observed at $\tilde{q} = \pi$ down
to the lowest energy probed, which demonstrates that the system
indeed remains gapless and critical at $H=8.7$ T.  The spectrum
has generally shifted to lower energy, and the strong ridge of scattering
with linear dispersion  near $\tilde{q} = \pi$ has a smaller slope
than at $H=0$, showing explicitly that the velocity of the elementary excitations
has decreased \cite{hammarprb1999}.  

There is a strong peak in $\tilde{{\mathcal I}}_{m}(\tilde{q},\omega)$
 centered at  $\tilde{q} = \pi$
at the field energy
 $\hbar\omega = g \mu_B H   \approx1.1$ meV ($g_c = 2.07$ \cite{McGregor76}).    
A weaker peak at this same energy can also be seen at $\tilde{q}=0$,
which corresponds to uniform spin precession.  
Moving away from  $\tilde{q} = \pi$, there is another 
ridge of scattering intensity, which decreases in energy
towards the expected field induced incommensurate wavevector
$\tilde{q}_{i,2} =0.7\pi$.
As shown in Fig.~\ref{fig:constQ}(e), the incommensurate mode is seen in
increased scattering at the lowest energies probed
at $\tilde{q}_{i,2}$ compared to zero field [Fig.~\ref{fig:constQ}(b)].  
Finally,  there is a mode that begins
at the field energy at $\tilde{q} = 0$, and decreases
in energy with increasing $\tilde{q}$, but which loses
intensity close to $\tilde{q}_{i,1} = 0.3\pi$, consistent with numerical
work \cite{lefmannprb1996}.

Using the Bethe Ansatz, M\"{u}ller {\it et al.} 
identified six classes of excitations out of the partially
magnetized ground state of an $N$-spin chain that can contribute
to $ {\mathcal S}(\tilde{q},\omega)$ \cite{muller1981}.  
Each class produces a continuum, and M\"{u}ller {\it et al.}
determined their approximate boundaries.  However,   three
of these are predicted to  dominate
$ {\mathcal S}(\tilde{q},\omega)$ as  $N \rightarrow \infty$, with
(adopting the notation of Ref.~\cite{muller1981},) 
class (ii) contributing to ${\mathcal{S}}_{zz}(\tilde{q},\omega)$,
class (iii) to ${\mathcal{S}}_{-+}(\tilde{q},\omega)$, and
class (vi) to ${\mathcal{S}}_{+-}(\tilde{q},\omega)$.
The boundaries of these three continua, 
${\cal E}_2$, ${\cal E}_3$, and ${\cal E}_6$, are shown as solid
lines in Fig.~\ref{fig:fieldcontour}. They
closely track the principal features of the data. 

We have calculated $ {\mathcal S}(\tilde{q},\omega)$ for chains of 
length $N=24$, $26$, and $28$, using  the Lanczos technique \cite{lefmannprb1996}.
Figure~\ref{fig:fieldcontour}(b) depicts the results 
 for $N=26$ and $m = 2/13$, corresponding to $H=8.67$ T for CuPzN
\cite{muller1981}. 
 Each eigenstate of the chain is marked by a circle with an
area  proportional to the corresponding contribution to $ {\mathcal S}(\tilde{q},\omega)$ 
\cite{lefmannprb1996}.
The circles are color-coded to indicate which component
of $ {\mathcal S}(\tilde{q},\omega)$ the state contributes to. 
Note  that  in our experimental geometry  we measure
  ${\mathcal {I}}_{m}(\tilde{q},\omega)\propto {\mathcal {S}}_{zz}+
  \frac{1}{4}({\mathcal {S}}_{+-}+{\mathcal {S}}_{-+})$.
The corresponding continuum boundaries  are also shown color-coded.
Some spectral weight for this finite chain is seen outside
of continua (ii), (iii) and (vi), and so
for reference the boundaries of the other three finite-$N$ 
continua, ${\cal E}_1$, ${\cal E}_44$, and ${\cal E}_5$, are included as dashed lines.
The finite-chain results reproduce all trends and features
of the measured intensity, and together with the continuum bounds,
suggest polarization assignments. 

A more direct comparison is achieved by combining the calculations in 
Fig.~\ref{fig:fieldcontour}(b)
with results for 
$N=24$, $m=1/6$ and $N=28$, $m=1/7$, which correspond
to $H=9.18$ T and 8.21 T for CuPzN, respectively, and greatly
increase the number of wavevectors sampled.
The results at each $N$ were smoothed
by convolving them with the response function of a finite-size system,
so that 
\begin{equation}
{\mathcal{S}}_{\alpha\beta}(\tilde{q}, \hbar\omega) = \sum_{N}\sum_{\tilde{q}_{N}}
{\mathcal{S}}_{\alpha\beta,N}(\tilde{q}_{N},
\hbar\omega)\left|\frac{\sin((\tilde{q}-\tilde{q}_{N})N/2)}{\sin((\tilde{q}-\tilde{q}_N)/2)}
\right|^{2}.
\end{equation}
The calculated results were then converted to
$\tilde{{\mathcal I}}_{m}(\tilde{q},\omega)$ \cite{hammarprb1999}, and
binned and averaged as described above. After multiplication
by an overall scale factor, we obtain the contour
plot shown in  Fig.~\ref{fig:fieldcontour}(c) which represents the measured 
intensity expected from an equal weighted ensemble of 24, 26, and 28 membered spin chains. 
While the relatively short chains yield stronger discrete energy bands than in 
the measurements, the simulation clearly captures the main features of the data. 

Karbach and M\"{u}ller have recently
identified a new class of quasiparticles for the 
partially magnetized S=1/2 chain.  These  ``psinons'' 
play a similar role in the spectrum  as do the spinons at $H=0$,
yielding a continuum for each component of $ {\mathcal S}(\tilde{q},\omega)$
\cite{karbach_prb_2000,Karbach02}, with
boundaries that are consistent with the approximate analytic
expressions that were used above.  
Karbach {\em et al.} have computed the psinon lineshapes  for $m = 0.25$, where
$\tilde{q}_{i,1} = \tilde{q}_{i,2} =\pi/2$.  This  occurs at
 $H = 11.9 $ T for CuPzN, but a qualitative comparison can still be made, as the
properties of the lineshapes should vary smoothly.  Of particular
interest is the peak at the upper boundary of the psinon continuum
for $ {\mathcal S}_{zz}(\tilde{q},\omega)$, which is due to a singularity in the 
psinon density of states.
This  may explain the
line of scattering intensity in Fig.~\ref{fig:fieldcontour}(a) that
tracks the top of the ${\cal E}_2$ continuum.
This is notably different from the zero-field case, where
the corresponding singularity in the spinon density of states is compensated by 
a vanishing matrix element,
and $ {\mathcal S}(\tilde{q},\omega)$ vanishes at the upper  boundary of 
the spinon continuum,
as seen in Fig.~\ref{fig:zerofieldcontour}.

Finally, we note that Fig.~\ref{fig:fieldcontour}(a) shows some evidence
of  weak scattering intensity for $\hbar\omega > 2 $ meV.  
This could possibly be due to  the presence
of short chains resulting from impurities, or  to higher-order
processes not included in the spinon/psinon picture.  However, we note
that our error bars are much larger  here
than at lower energy due to shorter counting times, (see Fig.~\ref{fig:constQ}), 
and so a definitive statement on the existence of excitations in this energy
range cannot be made at this time.

In summary, our experiments in CuPzN provide the first experimental example of an 
extended critical state in a quantum magnet. 
This detailed mapping of the
spin excitation spectrum in the spin-1/2 linear chain antiferromagnet verifies long-standing
predictions based on the Bethe Ansatz of a field driven and critical incommensurate state.
There is excellent agreement with finite chain calculations,
and good qualitative agreement with the lineshapes predicted
at higher fields based on novel ``psinon" quasiparticles.
Recent advances in Bethe Ansatz techniques
show promise for full calculations of the psinons' contribution
to $ {\mathcal S}(\tilde{q},\omega)$ \cite{Karbach02,Biegel02}.
A direct comparison of such calculations
to our data would clearly be very interesting.

This work was supported by NSF Grant No. DMR-0074571 and utilized
facilities supported by NIST and the NSF
under Agreement No. 9986442.  X-ray characterization was carried
out using facilities maintained by the JHU MRSEC 
under NSF Grant No. DMR-0080031.

\end{document}